\documentstyle[prc,aps]{revtex}
\draft

\newcommand{\be}{\begin{equation}}
\newcommand{\ee}{\end{equation}}
\newcommand{\ba}{\begin{eqnarray}}
\newcommand{\ea}{\end{eqnarray}}

\begin{document} 

\title{Elliptic Flow Fluctuations
}

\author{Stanis\l aw Mr\' owczy\' nski\footnote{Electronic address:
{\tt mrow@fuw.edu.pl}}}

\address{So\l tan Institute for Nuclear Studies \\
ul. Ho\.za 69, PL - 00-681 Warsaw, Poland \\
and Institute of Physics, \'Swi\c etokrzyska Academy \\
ul. \'Swi\c etokrzyska 15, PL - 25-406 Kielce, Poland}

\author{Edward V. Shuryak\footnote{Electronic address:
{\tt shuryak@dau.physics.sunysb.edu}}}

\address{Department of Physics and Astronomy \\
State University of New York at Stony Brook \\
Stony Brook, NY 11794-3800, USA}

\date{4-th March 2003}

\maketitle

\begin{abstract}

We suggest to perform systematic measurements of the elliptic flow 
fluctuations which are sensitive to the early stage dynamics of 
heavy-ion collisions at high-energies. Significant flow fluctuations 
are shown to be generated due to the formation of topological clusters
and development of the filamentation instability.  The statistical 
noise and hydrodynamic fluctuations are also estimated. 

\end{abstract}


\vspace{0.1in}


\section{Introduction}


A high-energy collision of heavy ions is often called the Little Bang
because of its similarity to the cosmological Big Bang. Both phenomena 
are violent explosions and both have attracted attention of 
experimentalists who have gathered unprecedented amount of data, 
limited basically by the data processing technology. The experiments
provided a lot of valuable information about the system's evolution. 
In particular, small variation in the temperature of background radiation 
have revealed mean dipole component, caused by the motion of the Solar 
System relative to the Big Bang heat bath. Tiny ($\sim 10^{-5}$) 
fluctuations on top of the dipole contribution, which have been
recently decomposed into angular harmonics with $l$ up to about 2000,
show peaks due to frozen sound exited 15 billions of years ago. A study 
of harmonic fluctuations in the Little Bang may possibly reveal something
interesting like frozen modulations as well.

One of the most spectacular experimental results obtained by now in 
relativistic heavy-ion collisions at RHIC is strong
elliptic flow quantified by mean value of the second
angular harmonics $v_2$ 
\cite{Ackermann:2000tr,Adler:2001nb,Adler:2002pu,Adcox:2002ms,Back:2002gz}. 
The phenomenon, which is sensitive to the collision early stage 
\cite{Sorge:1998mk} when the interaction zone is of the almond shape, is 
naturally explained within a hydrodynamics as a result of large density 
gradients 
\cite{Ollitrault:bk,Kolb:1999it,Kolb:2000sd,Teaney:1999gr,Teaney:2001av,Hirano:2001eu}.
Since, the hydrodynamic description is applicable for a system which 
is in local thermodynamic equilibrium, the large elliptic flow suggests 
a surprisingly short, below 1 ${\rm fm}/c$ \cite{Heinz:2001xi}, 
equilibration time which is difficult to reconcile with dynamic 
calculations, at least those performed within the perturbative QCD, 
see e.g. \cite{Heiselberg:1995sh}, where the early rapid expansion 
is closer to free streaming than to hydrodynamic evolution. We note here 
that the hydrodynamic model \cite{Heinz:2002rs}, which assumes the 
equilibration of only transverse degrees of freedom, has appeared rather
unsuccessful in describing the experimental data 
\cite{Ackermann:2000tr,Adler:2001nb,Adler:2002pu,Adcox:2002ms,Back:2002gz}. 
There have also been attempts \cite{Krasnitz:2002ng,Kovchegov:2002nf} 
to explain the large elliptic flow within models which do not invoke 
thermodynamic equilibrium. Finally, at large transverse momentum
$p_T=2-10$ GeV the magnitude of $v_2$ seem to be well described by
the surface emission model\cite{Shuryak:2001me}.

In such a situation, it is certainly desirable to look for experimental 
observables which can shed more light on the system dynamics at a 
collision early stage. We propose to go beyond measuring the {\em mean} 
elliptic flow magnitude, and study its (and higher harmonics, if it
ever be possible) {\em fluctuations  on the event-by-event 
basis\footnote{This is similar idea to the above mentioned measurement 
of the angular fluctuations of $T$. In cosmology, there is, of course, 
only one event but one can study angular fluctuations in various regions 
of the sky.}}. To be specific, we suggest to measure $v_2$ in every 
collision and then to analyze the variance of $v_2$. The first attempt 
of such an measurement has been undertaken in a very recent study by STAR 
Collaboration \cite{Adler:2002pu}. The result is however rather inconclusive. 
Our aim here is to motivate the work in order to improve experimental 
procedures. 

The elliptic flow fluctuations are shown to be sensitive even to somewhat 
exotic phenomena which have been argued to occur at the collision early 
stage. We consider here the filamentation instability 
\cite{Mrowczynski:qm,Mrowczynski:xv,Mrowczynski:1996vh} initiated due to 
the strong momentum anisotropy of the parton system, and the generation 
and subsequent explosions of the topological clusters \cite{Shuryak:2002an}. 
To detect these dynamical phenomena of interest one needs, however, a reliable 
estimate of the usual fluctuation. Therefore, a magnitude of the statistical 
noise is discussed in detail. We take into account the interference of various 
Fourier harmonics and finite resolution of the reaction plane reconstruction. 
We also study the fluctuations caused by the impact parameter and particle 
multiplicity variation.


\section{Formulation of the problem}
\label{formulation}


Two methods have been developed to quantify the ellipticity, as well
as higher harmonics, in the azimuthal angle distribution. One possibility 
is to work directly with the correlation functions of 2, 4, 6 or even 
more particles \cite{Borghini:2000sa,Borghini:2001vi}. Then, only the 
{\em relative} emission angles of particles matter and there is no need 
to determine a direction of the impact parameter $\psi_R$. Using the 
cumulants, one can partially eliminate detector effects, and consequently 
even detectors with relatively small acceptance can be used. The resulting 
Fourier coefficients include, however, not only the effects associated 
with  the ellipticity of events but are contaminated by any 2-, 4-, or, 
in general, $n$-body correlations caused by resonances, jets, quantum
statistics etc. However, it is hoped that these {\em nonflow}
correlations are dominated by the two-body effects and thus the 
genuine four-particle correlations provide rather clean information 
about the flow. Going to the six-particle correlations, the procedure
can be further improved.

In our considerations, we will refer to another method 
\cite{Voloshin:1994mz,Poskanzer:1998yz}, which was formulated earlier 
and is usually called a standard one. The method focuses on the angular
distributions relative to direction of the impact parameter. The 
experimental procedure splits in two steps which should be as independent 
from each other as possible. In the first step, one uses all available 
{\em multi-body} information about an event in order to determine the 
impact parameter direction $\psi_R$. In the second step, one constructs 
the distribution of the azimuthal angle relative to $\psi_R$ of `selected 
particles' and one computes the Fourier coefficients. The sets of particles 
used at these two steps are different from each other, and we will call 
their numbers as $M$ and $N$, respectively. In order to reduce non-flow 
correlations, the particles of both sets (subevents) are usually separated 
by a rapidity gap. They are still correlated by the flow because the 
direction of the impact parameter $\psi_R$ is a global feature of an 
event, like the magnitude of the impact parameter itself\footnote{In 
principle, the impact parametr magnitude can be reconstructed from 
2-, 4-, 6-, ..., $n$-particle correlators. However, it would be very 
difficult and presumably rather inaccurate method. Using total 
multiplicity or forward calorimeter signal does the job very well.}. 
In practice, it is desired to use the cumulant and standard method 
simultaneously. Comparing the results, one can eventually separate 
the `flow' (global correlation between all secondaries) from `local 
correlations' involving a few particles only. Such a comparison made 
by STAR \cite{Adler:2002pu} has shown that unless one goes to high 
$p_T$ or very central collisions, the flow dominates and 2-body 
correlations contribute to $v_2$ obtained from the standard method 
only at the 10-15 \% level.

Since the one-particle distribution in a single event can be 
written as
\be \label{Fourier}
P(\phi) = {1 \over 2\pi} \;
\Big[1 + 2\sum_{n=1}^{\infty} 
v_n {\rm cos}\big(n(\phi- \psi_R)\big) \Big] 
\Theta(\phi) \, \Theta(2\pi - \phi) \;,
\ee
the $n-$th Fourier amplitude is determined as
$$
v_n = \overline{{\rm cos}\big(n(\phi - \psi_R)\big)} \;,
$$
with $\overline{\cdots}$ denoting averaging over one-particle 
distribution in a single event.

The reaction plane is never reconstructed precisely and the real 
reaction plane angle $\psi_R$ deviates from the estimated angle $\psi_E$. 
One observes that 
$$
v_n = {1 \over R_n} \; \overline{{\rm cos}\big( n(\phi -\psi_E)\big)} \;,
$$
where $R_n \equiv {\rm cos}\big( n(\psi_R -\psi_E)\big)$ is the 
reaction plane resolution factor.

Let us now think about the ensemble of events with every event representing
a single nucleus-nucleus collision. The angular harmonics $v_n$ are measured
for each event. It should be stressed that this is not only the angle 
$\psi_R$ (and $\psi_E$) which varies form event to event but the amplitudes 
of Fourier harmonics can also vary due to dynamical reasons. According 
to \cite{Poskanzer:1998yz}, the average over events of the harmonic's 
amplitude is not defined as 
\be \label{def-ave1-v2} 
\langle v_n \rangle \buildrel \rm def \over = 
\Bigg\langle 
{\overline{{\rm cos}\big( n(\phi -\psi_E)\big)} \over R_n} \Bigg\rangle 
\ee
but 
\be \label{def-ave2-v2} 
\langle v_n \rangle \buildrel \rm def \over = 
{\Big\langle  \overline{{\rm cos}\big( n(\phi -\psi_E)\big)} \Big\rangle 
\over \langle R_n \rangle } \;,
\ee
where $\langle \cdots \rangle$ denotes averaging over events.
Since the procedure of determining the angle $\psi_E$ is arranged 
to be maximally independent from that of computing of 
$\Big\langle \overline{{\rm cos}\big( n(\phi -\psi_E)\big)}\Big\rangle$, 
it is expected that the event averaging of $R_n$ and of 
$\overline{{\rm cos}\big( n(\phi -\psi_E)\big)}$ are 
independent from each other. If so,
$$
\Bigg\langle 
{\overline{{\rm cos}\big( n(\phi -\psi_E)\big)} \over R_n }  
\Bigg\rangle = 
\bigg\langle {1 \over R_n} \bigg\rangle 
\Big\langle \overline{{\rm cos}\big( n(\phi -\psi_E)\big)} 
\Big\rangle \approx
{\Big\langle \overline{{\rm cos}\big( n(\phi -\psi_E)\big)} \Big\rangle 
\over
\langle R_n \rangle } \;,
$$
where the approximate equality holds if the resolution factor 
${\rm cos}\big( n(\psi_R -\psi_E)\big)$ does not much vary from event 
to event. Thus, the definitions (\ref{def-ave1-v2}, \ref{def-ave2-v2}) 
are approximately equivalent to each other, provided $\psi_R$ is
reconstructed sufficiently well (which requires $M \gg 1$).

While the definition (\ref{def-ave1-v2}) can be uniquely 
extended to the second moment, there is an ambiguity how
to generalize the definition (\ref{def-ave2-v2}). Therefore,
we define 
$$
\langle v_n^2 \rangle \buildrel \rm def \over = 
{ 1 \over \langle R_n \rangle^2 } 
\Big\langle \overline{{\rm cos}\big( n(\phi -\psi_E)\big)}^2 
\Big\rangle \;,
$$
and the fluctuations as
\be \label{def-var-vn}
{\sf Var}(v_n) \buildrel \rm def \over = 
\langle v_n^2 \rangle - \langle v_n \rangle^2 
= { 1 \over \langle R_n \rangle^2 } \Big(
\Big\langle \overline{{\rm cos}\big( n(\phi -\psi_E)\big)}^2 
\Big\rangle 
- \Big\langle \overline{{\rm cos}\big( n(\phi -\psi_E)\big)}
\Big\rangle^2 \Big) \;,
\ee
where $\langle R_n \rangle^2$ enters as a multiplicative factor. 
However, $R_n$ generates event-by-event fluctuations of observed 
$v_n$. To see the effect, let us consider the single-particle 
azimuthal distribution in a given event of the form (\ref{Fourier}) 
with the amplitudes $v_n$ being exactly the same in {\em all} events.  
Then, $\langle v_n \rangle = v_n$ but
$$
{\sf Var}(v_n) = {\langle R_n^2 \rangle - \langle R_n \rangle^2
\over \langle R_n \rangle^2} \; v_n^2 \;.
$$
Thus, the fluctuations of $R_n$ contribute to ${\sf Var}(v_n)$.
We will often use the symbol $\delta v_n \equiv \sqrt{{\sf Var}(v_n)}$.

In the following sections, we will focus our attention
on the second harmonics and consider several sources of 
the $v_2$ fluctuations.


\section{Statistical Noise}
\label{noise}


We start our discussions of the fluctuations of $v_2$ with those 
caused by the finite number $N$ of particles which are used at the 
step 2 of the standard method when the Fourier amplitudes are
determined. We assume here that $v_n$ do {\em not} change from 
event to event. We also assume that the only correlations in 
the system are those due to the flow. Then, the azimuthal distribution 
of $N$ particles is a product of $N$ single particle distributions. 
Namely,
$$
P_N(\phi_1,\phi_2, \cdots, \phi_N) = {\cal P}_N
P(\phi_1) \: P(\phi_2) \; \cdots \; P(\phi_N) \;,
$$
where ${\cal P}_N$ is the multiplicity distribution while all 
distributions $P(\phi_i)$ are given by Eq.~(\ref{Fourier}).
The single particle distributions $P(\phi_i)$ are correlated to
each other because of the common angle $\psi_R$.

In a single event, the ellipticity is found as
$$
v_2 = {1 \over R_2} \; { 1 \over N}
\sum_{i=1}^N {\rm cos}\big(2(\phi_i -\psi_E)\big)
$$
where $\phi_i$ is the azimuthal angle of $i-$th particle and
$N$ is the event's multiplicity. According to the definition 
(\ref{def-ave2-v2}) the ensemble average of $v_2$ then equals
\ba \label{ave-v2-noise}
\langle v_2 \rangle &=& 
{1\over \langle R_2 \rangle} 
\bigg\langle 
{1 \over N} \sum_{i=1}^N {\rm cos}\big(2(\phi_i -\psi_E)\big)
\bigg\rangle \\ [2mm] \nonumber 
&=& \sum_{N=1}^{\infty} {\cal P}_N \, {1 \over N}
\int_0^{2\pi}d\phi_1 P(\phi_1) \int_0^{2\pi} d\phi_2 P(\phi_2) \cdots 
\int_0^{2\pi}d\phi_N P(\phi_N ) 
\sum_{i=1}^N {\rm cos}\big(2(\phi_i -\psi_E)\big)  = v_2 \;.
\ea
We note that the event-by-event averaging of $R_2$ and of
${\rm cos}\big(2(\phi_i -\psi_E)\big)$ are assumed to be
independent from each other.

The second moment is
\ba \label{var-v2-noise}
\langle v_2^2 \rangle &=& 
{1\over \langle R_2 \rangle^2} \bigg\langle \Big(
{1 \over N} \sum_{i=1}^N {\rm cos}\big(2(\phi_i -\psi_E)\big)
\Big)^2 \bigg\rangle 
\\ [2mm] \nonumber
&=& {1\over \langle R_2 \rangle^2} 
\sum_{N=1}^{\infty} {\cal P}_N \, {1 \over N^2}
\int_0^{2\pi}d\phi_1 P(\phi_1) \int_0^{2\pi} d\phi_2 P(\phi_2) \cdots 
\int_0^{2\pi}d\phi_N P(\phi_N ) 
\Big(\sum_{i=1}^N {\rm cos}\big(2(\phi_i - \psi_E)\big)\Big)^2  
\\ [2mm] \nonumber
&=& 
{1\over \langle R_2 \rangle^2} \bigg[ \Big( {1\over 2} 
+ {1 \over 2}\: v_4 \: (2\langle R_2^2 \rangle - 1) \Big)
\Big\langle {1\over N} \Big\rangle 
+ v_2^2 \langle R_2^2 \rangle 
\Big\langle{N - 1\over N}\Big\rangle \bigg] \;.
\ea

It has been found in Au-Au collisions at RHIC \cite{Adler:2002pu} 
that $\langle v_4 \rangle \ll \langle v_2 \rangle$  while 
$\langle v_2 \rangle$ reaches the value of about 0.07 for rather 
peripheral collisions. Taking these numbers into account,
we estimate the fluctuations of $v_2$ as
\be \label{noise2}
{\sf Var}(v_2) = {1\over 2 \langle R_2 \rangle^2 \langle N \rangle} 
+ \langle v_2 \rangle^2 \;
{\langle R_2^2 \rangle - \langle R_2 \rangle^2 \over 
\langle R_2 \rangle^2} \;,
\ee
where we have also assumed that $\langle N \rangle \gg 1$ and
that the multiplicity fluctuations are small. 

The second term in r.h.s of Eq.~(\ref{noise2}) depends on the number 
of particles $M$ used to determine the impact parameter direction.
Since $M$ is usually  rather large, we assume in accord with 
\cite{Adler:2002pu} that $R_2$ does not much deviate from unity. 
Then, as argued in \cite{Voloshin:1994mz,Poskanzer:1998yz}, we have
\be \label{ave-R}
\langle R_2 \rangle = 
\langle {\rm cos}\big( 2(\psi_R -\psi_E)\big) \rangle 
\approx 1 - \langle (\psi_R -\psi_E)^2 \rangle 
\approx 1 - {a \over 2 \langle M \rangle}
\ee
where the parameter $a$ depends on the type of weights which are applied. 
An actual value of $a$ is irrelevant for our considerations. Using the 
arguments which lead us to the result (\ref{ave-R}), one finds
$$
\langle R_2^2 \rangle = 
\langle {\rm cos}^2\big( 2(\psi_R -\psi_E)\big) \rangle 
\approx 1 - 2 \langle (\psi_R -\psi_E)^2 \rangle  
\approx 1 - {a \over \langle M \rangle } \;.
$$
Thus, $\langle R_2^2 \rangle - \langle R_2 \rangle^2 \sim 
\langle M \rangle^{-2}$. Since the number of particles used to 
determine the reaction plane is larger or at least similar to that 
which is involved in finding $v_2$, we conclude that the second term 
in r.h.s of Eq.~(\ref{noise2}) can be neglected. Thus, we finally 
estimate the statistical noise as
\be \label{noise1}
\delta v_2 = {1\over \langle R_2 \rangle \sqrt{2 \langle N \rangle }}\;.
\ee

As an extra check, we have performed a Monte Carlo simulation of fake events 
with $N$ particles generated according to 
$P(\phi) \sim \big(1 + 2v_2 {\rm cos}(2(\phi- \psi_R)) \big)$. The obtained
variation of $v_2$ is, of course, in full agreement with the expression 
given above.

In the subsequent sections, we discuss physical phenomena which lead to the 
fluctuations of $v_2$ different than those described by Eq.~(\ref{noise1}),
i.e. originating from true event-by-event fluctuations of flow.


\section{Impact parameter and multiplicity fluctuations}
\label{hydro}


As already noted, the observed elliptic flow is naturally described in 
the hydrodynamic model 
\cite{Ollitrault:bk,Kolb:1999it,Kolb:2000sd,Teaney:1999gr,Teaney:2001av,Hirano:2001eu}.
Therefore, are going to discuss here how large are the fluctuations of 
$v_2$ within the hydrodynamics. One should distinguish the fluctuations 
due to the varying impact parameter and those due to the thermodynamic 
fluctuations at fixed collision geometry. We start with the former ones.

As well known, $\langle v_2 \rangle$ strongly depends on the collision
impact parameter $b$. In the case of Au-Au collisions at RHIC at 
$\sqrt{s_{NN}} = 130$ GeV, the dependence has been parameterized 
\cite{Adler:2002pu} as
\be \label{v2-b}
\langle v_2 \rangle = a_1b + a_2b^2 + a_3^3 + a_4b^4
+a_5b^5 + a_6 b^6 \;,
\ee
where $b$ is measured in fm and $a_1 = - 3.94 \cdot 10^{-4}$,
$a_2 = 2.1 \cdot 10^{-3}$, $a_3 = - 7.06 \cdot 10^{-5}$,
$a_4 = - 3.2 \cdot 10^{-5}$, $a_5 = 3.58 \cdot 10^{-6}$,
and $a_6 = - 1.17 \cdot 10^{-7}$. The parameterization assumes
that $\langle v_2 \rangle$ vanishes for $b=0$ and for
$b > b_{\rm max} = 14.7$ fm.

The $v_2$ fluctuations  due to the varying impact parameter 
can be estimated by the formula
$$
\delta v_2 = 
{d \langle v_2 \rangle \over d\,b}\; \delta b \;.
$$
The impact parameter fluctuations are, in principle, measurable through 
the observation of multiplicity of participating nucleons $N_p$ which
in turn is controlled by the collision trigger conditions. Then, $\delta b$ 
can be recalculated into $\delta N_p$. Adopting the linear dependence 
$$
N_p = 2Z \Big( 1 - { b \over b_{\rm max}} \Big) \;,
$$
where $Z = 79$ is the number of protons in a gold nucleus, one gets
the formula
\be \label{v2fluc-N_p}
\delta v_2 = 
\big(a_1 + 2\, a_2b + 3\, a_3^2 + 4\, a_4b^3 + 5\, a_5b^4 
+ 6\, a_6 b^5 \big)\;{ b_{\rm max} \over 2Z} \; \delta N_p \;,
\ee
We note that for $b \approx 10$ fm where the flow is maximal
the $v_2$ fluctuations due to the impact parameter variation 
vanish because the derivative $d \langle v_2 \rangle /db$ is 
then zero. For $b = 5$ fm where $\langle v_2 \rangle \approx 0.03$, 
Eq.~(\ref{v2fluc-N_p}) gives 
$\delta v_2 \approx 8 \cdot 10^{-4}\: \delta N_p$ which should
be compared to the statistical noise (\ref{noise1}). For $b=5$ fm,
$\delta N_p = 30$ and $\langle N \rangle = 500$, the magnitude
of the $v_2$ fluctuations caused by the impact parameter variation 
is approximately equal to that of statistical noise. Thus, not only
the statistical noise but also the centrality fluctuations must be 
subtracted from the measured $v_2$ fluctuations to observe the 
dynamical fluctuations of interest. 

Let us now consider the fluctuations of $v_2$ due to the variation 
of thermodynamic parameters. The most important are presumably the 
multiplicity fluctuations. Here, we present some general formulas, 
considering an example of non-statistical fluctuations of multiplicity 
in the next section. (We follow a similar analysis \cite{Shuryak:2002an} 
of the mean $p_T$ event-by-event fluctuations due to radial flow 
fluctuations.) We assume here that the multiplicity of produced
particles is not directly used to determine the collision centrality.
In such a case the predicted $v_2$ fluctuations could be significantly
reduced. 

The fluctuations of $v_2$ can be estimated as
$$
\delta v_2 = 
{d \langle v_2 \rangle \over d\langle N \rangle }\; \delta N \;,
$$
which can be rewritten in the form 
\be \label{hydro-1}
{\delta v_2 \over \langle v_2 \rangle }
={\delta N \over \langle N \rangle } P_h \;,
\ee
where the index $P_h$ (the effective power) is
$$
P_h \equiv 
{d {\rm ln} \langle v_2 \rangle \over d {\rm ln} \langle N \rangle} 
= {\langle N \rangle \over \langle v_2 \rangle} \;
{d \langle v_2 \rangle \over d \langle N \rangle} \;,
$$
with $h$ denoting a hadron species used to determine $v_2$. We note 
that STAR collaboration has already reported the data on $v_2$ for 
$\pi,K,K_s,p,\Lambda$ \cite{Adler:2001nb}. The first {\it stochastic} 
factor in Eq.~(\ref{hydro-1}) is the relative multiplicity fluctuation 
which drives the fluctuations of $v_2$, while the second {\it dynamical} 
factor $P_h$ shows how a change in entropy transfers into $v_2$. 
$P_h$ is obviously different for various secondary hadron species
which can be used to test the idea further.

Assuming the poissonian character of multiplicity fluctuations,
Eq.~(\ref{hydro-1}) can be rewritten as
\be \label{hydro-2}
\delta v_2 = {\langle v_2 \rangle \over \sqrt{N}} \; P_h \;.
\ee

The value of the index $P_h$ can be estimated within the hydrodynamics. 
The calculations presented in \cite{Teaney:2001av} for EoS LH8 show 
that changing $dN/dy$ from 200 to 400 leads to the increase of $v_2$ 
for pions from 0.028 to about 0.04 in a good agreement with NA49 and 
STAR data, see Fig.~24 in \cite{Teaney:2001av}. Reading a logarithmic 
slope from that figure, we find $P_\pi \approx 0.4$, which will be 
used below\footnote{P.~Kolb was kind enough to provide the results for 
pions and nucleons from his version of hydrodynamics, and the indices
turned out to be three times smaller, $P_\pi\approx 0.12, 
P_p\approx 0.13$. These number, however, are affected by some artifacts 
of the freeze-out approximations used, especially for low (SPS) energies. 
As a result, there is a non-monotonous dependence of $v_2$ vs. $dN/dy$ 
with a minimum, reducing the index. We note that a compilation of the 
AGS-SPS-RHIC data show a monotonous rise, as found in \cite{Teaney:2001av}.}.

Comparing Eqs.~(\ref{noise1}) and (\ref{hydro-2}), one finds that the 
ratio of the hydrodynamic fluctuations to the statistical noise is 
$\sqrt{2}\, \langle R_2 \rangle \langle v_2 \rangle \, P_h$ which for 
$\langle R_2 \rangle = 0.6$, $\langle v_2 \rangle = 0.07$ 
\cite{Adler:2002pu} and $P_h = 0.4$ is 0.02. The effect is indeed 
rather small. However, as discussed in Sec. \ref{noise}, the magnitude 
of the statistical noise can be well controlled, and consequently, the 
hydrodynamic fluctuations seem to be detectable. 


\section{Fluctuations Induced by  Cluster Formation}
\label{clusters}


In general, cluster production induce  event-by-event fluctuations 
of local multiplicity or $dN/dy$ larger than pure statistical noise, 
simply because the number of clusters is smaller than the number of 
particles. The estimates, which we will give, follow ordinary 
statistical arguments, assuming that the cluster production happens 
in statistically independent way. This is justified by the observation 
that they all appear at different locations in the transverse plane, 
and also that they depend on specific hadronic and vacuum configurations 
at the moment of the collision. The original presentation of the idea 
that clusters should lead to observable event-by-event fluctuations 
of flows has been made by one of us at the 2001 CERN workshop
\cite{fluct_hif}. 

Experimentally, existence of strong clustering of produced secondaries 
in p-p collisions has been known since 2-body correlation function 
was measured at ISR long time ago. To our knowledge, however, precise
properties of such clusters have been never really explained or well 
quantified. As a relatively recent example of the cluster study, we 
refer to Fermilab experiment \cite{Alexopoulos:ft} where high multiplicity 
${\rm \bar p}$-p collisions have been analyzed with the conclusion that 
the average charged multiplicity per cluster is about 4. A single 
isolated cluster is produced in the so-called Pomeron-Pomeron process. 
Another example is provided by a recent analysis of the old UA8 data 
\cite{Brandt:2002qr}, showing production of clusters of 3-5 GeV mass. 
It was further shown that such clusters, with mass up to 5 GeV, decay 
isotropically in their rest frame\footnote{Unfortunately, the UA2 detector, 
which was used to collect the data, was just a simple calorimeter, and 
we do not know anything about the structure of these clusters or even 
mean multiplicities. RHIC detectors and especially STAR can do a lot 
of clarification in p-p mode, provided proper triggers are implemented.}. 
Understanding of such clusters is very important to clarify a long-standing 
problem of `soft Pomeron' dynamics.

As a theoretical motivation, we suggest topological cluster 
formation in heavy-ion collisions. Inhomogeneous structure of 
the QCD vacuum, with relatively dilute gas of instantons, results 
in also dilute set of topological clusters arising at the collision 
early stage when the system is promptly excited, from virtual to 
real classical fields. For more discussion of these ideas, specific 
formulas and original references, see recent paper by one 
of us \cite{Shuryak:2002an} and some subsequent works 
\cite{Shuryak:2002qz,Janik:2002nk} where the cluster production and 
decay into gluons and quarks is discussed. For the purpose of this 
paper, it is enough to know that such a cluster, a QCD sphaleron, is 
like a heavy resonance which is expected to decay into about 3 gluons 
and 6 quarks and antiquarks. In p-p those should hadronize into 
specific final states, while in heavy ion case these partons are 
absorbed by the fireball and simply increase the local entropy
density. This should cause event-by-event fluctuations of radial 
flow \cite{Shuryak:2002an} as well as of elliptic flow we discuss
in this note.

Let us now quantify particle number fluctuations caused by the 
cluster formation. As previously, $N$ denotes particle multiplicity 
in a given $p_T$ and $y$ window and $N_{\rm cl}$ is the number of 
hadrons which can be attributed to the cluster decays. With $N_0$ 
we denote the hadron multiplicity from all sources different than the 
topological clusters. Then, the relative fluctuation of the particle 
multiplicity can be written as
$$
{\delta N \over \langle N \rangle} = 
{\langle N_0 \rangle \over \langle N \rangle} 
{\delta N_0 \over \langle N_0 \rangle} +
{\langle N_{\rm cl} \rangle \over \langle N \rangle} 
{\delta N_{\rm cl} \over \langle N_{\rm cl} \rangle} \;.
$$
Now, we assume that the fluctuations of both the cluster number 
$n_{\rm cl}$ and $N_0$ are poissonian, i.e.
$$
{\delta n_{\rm cl} \over \langle n_{\rm cl} \rangle} =
{1 \over \sqrt{\langle n_{\rm cl}\rangle}} \;,\;\;\;\;\;\;\;\;\;\;
{\delta N_0 \over \langle N_0 \rangle} = 
{1 \over \sqrt{\langle N_0 \rangle}} \;,
$$ 
and that every cluster provides $k$ hadrons 
($ N_{\rm cl} = k \, n_{\rm cl}$). Then, the relative fluctuation of 
the hadron number is
\be \label{N-fluc}
{\delta N \over \langle N \rangle} = 
{1 \over \sqrt{\langle N \rangle}} \; 
\big(\sqrt{1 - f} + \sqrt{f k} \;\big) \;,
\ee
where $f \equiv \langle N_{\rm cl} \rangle / \langle N \rangle$ denotes 
the fraction of final state hadrons produced due to the cluster
formation.

Assuming that the cluster production alone is completely responsible 
for the p-p cross growth with the collision energy, one obtains the 
{\em upper bound}\footnote{This is an upper bound because nuclear
modification of the structure functions is ignored. The realistic 
number is presumably factor 2 or so smaller.} on cluster production 
in Au-Au \cite{Ostrovsky:2002cg}. Adopting the scaling from p-p 
to Au-Au with the number of hard collisions, it was estimated in 
\cite{Ostrovsky:2002cg} that up to roughly 70 clusters per unit rapidity, 
$dn_{\rm cl}/dy \approx 70$, can be produced around $y=0$ in central 
Au-Au collisions at RHIC. This estimate, in turn, leads to an (upper 
limit) on cluster-related entropy of about {\em half} of the total value, 
and consequently of about {\rm half} of the total multiplicity ($f \le 0.5$). 
Keeping in mind that $d N/dy \approx 550$, one gets $k \approx 4$. 
Inserting these numbers into Eq.~(\ref{N-fluc}), we find that the formation 
and subsequent decays of clusters can (maximally) double the multiplicity 
fluctuations when compared to the poissonian fluctuations.

Using Eq.~(\ref{hydro-1}), one immediately translates the multiplicity
fluctuations into the fluctuations of $v_2$. Since, the clusters can 
even double $\delta N /\langle N \rangle$, the same holds for 
$\delta v_2 /\langle v_2 \rangle$. Once we have concluded Sec. \ref{hydro}
that the hydrodynamic fluctuations seem to be measurable, we claim here 
that there is a chance to detect the fluctuation growth due to the 
cluster formation. However, it should stressed that the production
and subsequent decay of the clusters can be observed directly studying
the multiplicity fluctuations.


\section{Fluctuations induced by Filamentation instability}
\label{filamen}


When the momentum distribution of partons is strongly elongated in one 
direction, say along the $z$ i.e. beam axis, the neutral system has 
a tendency to split into the filaments along $z$ with the current flowing 
in the opposite directions in neighboring filaments. The reason is as 
follows: once the currents in the system occur, they generate the 
(chromo-)magnetic field, oscillating in the direction perpendicular 
to the beam axis, and the Lorentz force acts back on the charges which 
form the currents. It appears that the currents get focused and the 
current magnitude grows. This is the filamentation instability 
\cite{Wei59} which have been studied in the context of ultrarelativistic 
heavy-ion collisions in 
\cite{Mrowczynski:qm,Mrowczynski:xv,Mrowczynski:1996vh}. 

The breakdown of the azimuthal symmetry of the system due to the 
instability development gives a chance to observe it experimentally. 
It has been argued 
\cite{Mrowczynski:qm,Mrowczynski:xv,Mrowczynski:1996vh} that the instability 
growth leads to the energy transport along the wave vector which coincides 
with the Poynting vector of the generated chromodynamic field. Consequently, 
one expects significant a variation of the transverse energy as a function 
of the azimuthal angle. 

Here we point another, presumably more realistic, possibility to detect the 
color filamentation. When the instability grows the trajectories of charge 
particles are focused in the centers of the filaments. Therefore, according 
to the Liouville theorem the distribution of the momentum perpendicular to 
the filaments, say along the $x-$axis, has to expand to conserve the phase 
space volume. The quantum mechanical counterpart of the argument relies on 
the uncertainty relation: once the particles are localized within the 
filaments their transverse momentum has to widen to the inverse filament 
thickness multiplied by $\hbar$. Below, we quantify this quantum mechanical 
reasoning.

It should be clearly stated that the collective motion caused by
the instability development is not correlated with the reaction plane
and it has nothing to do with the hydrodynamic flow.  As such it would
be called a `non-flow' effect. However, the filamentation generates 
a finite value of $v_2$ and it contributes to its fluctuations.
So, let us consider the phenomenon in more detail. 

Let the wave vector of the filamentation mode ($k$) be oriented along the 
$x-$axis. Then, the single particle wave function describing the transverse 
degrees of freedom is the form 
\be \label{wave-x}
\psi (x,y) \sim {\rm exp}\Big[-{x^2 + y^2 \over 4 R^2} \Big]
\; {\rm cos}(kx + \alpha) \;,
\ee
where $R$ is the system transverse radius. To simplify further analysis 
we put the phase $\alpha$ equal to zero. Then, as we will see, odd harmonics 
of the azimuthal distribution vanish due to the mirror symmetry of the
wave function (\ref{wave-x}). Performing the Fourier transform of 
(\ref{wave-x}), one gets the momentum distribution as
\be \label{p-distribution}
P (p_x,p_y) = |\widetilde{\psi} (p_x,p_y) |^2
\sim {\rm exp} \big[-2 R^2(p_x^2 + p_y^2)\big]
\Big[ {\rm exp} \big(4 R^2p_x k\big) +
{\rm exp}\big(-4 R^2p_x k \big) + 2 \Big] \;.
\ee
Since $p_x = p \,{\rm cos}\phi$ and $p_y = p \, {\rm sin}\phi$, the 
distribution (\ref{p-distribution}) provides the azimuthal distribution 
of the form
\be \label{phi-dist-fil1}
P(\phi) = \int_0^{\infty}dp \, p \, 
P(p \, {\rm cos}\phi , p \, {\rm sin}\phi) 
\sim \int_0^{\infty}dp \;p \; e^{-2R^2p^2} \Big[
e^{ 4R^2pk\,{\rm cos}\phi } +
e^{-4R^2pk\,{\rm cos}\phi } + 2 \Big] \;.
\ee
Neglecting the jacobian $p$ in Eq.~(\ref{phi-dist-fil1}), the integral 
over momentum can be performed analytically and the result reads
\be \label{phi-dist-fil2}
P(\phi) \sim \Big[
e^{ 2R^2k^2\,{\rm cos}^2\phi } + 1 \Big] \;.
\ee
The distribution (\ref{phi-dist-fil2}) gets a particularly simple form 
when the filament thickness is much smaller than the system size. Then, 
$R \, k \gg 1$ and 
\be \label{phi-dist-fil3}
P(\phi) = {1 \over 2(\pi -1)} \Big[1 -
\delta (\phi -\pi/2) - \delta (\phi -3\pi/2) \Big] \;.
\ee

Using the distribution (\ref{phi-dist-fil3}), one finds
\be \label{fil-v2}
v_2 = {1 \over \pi -1} \;.
\ee
The value of $v_2$ is rather large. A realistic value of $v_2$
is presumably significantly smaller because not all particles 
produced in a given event would participate in a collective motion 
caused the instability development. The collective motion should
be also convoluted with the thermal one. Thus, the effect of 
filamentation must be diluted. An appearance of the instability 
in the system is not a deterministic but a random process. Therefore, 
we expect that there are collisions with and without the instability. Consequently, $v_2$ varies between zero and maximal value (\ref{fil-v2}). 
Thus, we expect large fluctuations of $v_2$.  

It should be also stressed here that there are specific distinctive 
features of the collective flow and the flow fluctuations due to the 
filamentation. First of all we note that in contrast to hydrodynamically 
generated $v_2$, the flow caused by the instability development does 
not vanish at zero impact parameter. Thus, one should look for filamentation 
in maximally central collisions. It is also expected that particles
with small $p_T$ are particularly sensitive the collective motion
of interest.


\section{Final remarks}


The aim of this paper is to advocate usefulness of the flow fluctuation 
analysis in revealing of the early stage dynamics of heavy-ion collisions. 
We have shown that even rather exotic phenomena can be studied in this
way. Those presented should be, of course, treated only as examples 
motivating the measurements, and simple order-of-magnitude estimates.
 
Since $v_2$ is experimentally determined on the event-by-event basis
anyway, the proposed fluctuation measurement presumably does not require
much additional efforts. However, an accuracy of the measurements should
be improved. Studying the flow fluctuations as a function of particle 
multiplicity one can check whether $\delta v_2$ scales like 
$1/\sqrt{\langle N \rangle}$, which is a characteristic feature of 
statistical noise. If not we deal with nontrivial dynamical 
fluctuations. However, before such a conclusion is achieved one has 
to properly subtract the fluctuations due to the impact parameter 
variation. We note that these fluctuations can be constrained by 
the collision trigger condition and that the fluctuations due to 
impact parameter vanish around the maximal flow where 
$d \langle v_2 \rangle /db = 0$.

To disentangle various fluctuation sources, the data should be analyzed 
in a broad interval of impact parameters and varying acceptance windows. 
Fortunately, the mechanisms of interest contribute differently to 
$\delta v_2$. In particular, the cluster effect is expected to be the 
largest for most peripheral events, while that of the filamentation for 
the most central ones. It would be also desirable to study elliptic 
flow fluctuations simultaneously with fluctuations of particle 
multiplicity and other collision characteristics. 

At the end, we mention one more supplementary method 
\cite{Mrowczynski:1999vi} to study the azimuthal fluctuations which 
seem to be particularly useful to detect non-flow correlations. The 
method, which uses the so-called $\Phi-$measure of fluctuations 
\cite{Gazdzicki:ri}, does not require the reaction plane reconstruction 
and can be rather easily applied to experimental data. The measure is 
sensitive to various sources of dynamical correlations and the integrated 
information provided by $\Phi$ can be combined with that offered by the 
Fourier analysis \cite{Voloshin:1994mz,Poskanzer:1998yz}.  Since all 
Fourier harmonics contribute to $\Phi$ one can check whether the measured 
harmonics saturate the observed value of $\Phi$.

\section{Acknowledgements} 

Our thanks go to Peter Kolb who shared with us his unpublished results 
of hydrodynamic calculations. We are also indebted to Jean-Yves 
Ollitrault and Art Poskanzer for very useful correspondence. St. M. is 
grateful to Department of Physics and Astronomy of State University of 
New York at Stony Brook, where this project was initiated, for kind 
hospitality. This work was partially supported by the US Department of 
Energy and Polish Committee of Scientific Research under grants 
DE-FG02-88ER40388 and 2P03B04123, respectively.


\begin{thebibliography}{99}

\bibitem{Ackermann:2000tr}
K.~H.~Ackermann {\it et al.}  [STAR Collaboration],
Phys.\ Rev.\ Lett.\  {\bf 86}, 402 (2001).

\bibitem{Adler:2001nb}
C.~Adler {\it et al.}  [STAR Collaboration],
Phys.\ Rev.\ Lett.\  {\bf 87}, 182301 (2001).

\bibitem{Adler:2002pu}
C.~Adler {\it et al.}  [STAR Collaboration],
Phys.\ Rev.\ C {\bf 66}, 034904 (2002).

\bibitem{Adcox:2002ms}
K.~Adcox  [PHENIX Collaboration],
Phys.\ Rev.\ Lett.\  {\bf 89}, 212301 (2002).

\bibitem{Back:2002gz}
B.~B.~Back {\it et al.}  [PHOBOS Collaboration],
Phys.\ Rev.\ Lett.\  {\bf 89}, 222301 (2002).

\bibitem{Sorge:1998mk}
H.~Sorge,
Phys.\ Rev.\ Lett.\  {\bf 82}, 2048 (1999).

\bibitem{Ollitrault:bk}
J.~Y.~Ollitrault,
Phys.\ Rev.\ D {\bf 46}, 229 (1992).

\bibitem{Kolb:1999it}
P.~F.~Kolb, J.~Sollfrank and U.~W.~Heinz,
Phys.\ Lett.\ B {\bf 459}, 667 (1999).

\bibitem{Kolb:2000sd}
P.~F.~Kolb, J.~Sollfrank and U.~W.~Heinz,
Phys.\ Rev.\ C {\bf 62}, 054909 (2000).

\bibitem{Teaney:1999gr}
D.~Teaney and E.~V.~Shuryak,
Phys.\ Rev.\ Lett.\  {\bf 83}, 4951 (1999).

\bibitem{Teaney:2001av}
D.~Teaney, J.~Lauret and E.~V.~Shuryak,
arXiv:nucl-th/0110037.

\bibitem{Hirano:2001eu}
T.~Hirano,
Phys.\ Rev.\ C {\bf 65}, 011901 (2002).

\bibitem{Heinz:2001xi}
U.~W.~Heinz and P.~F.~Kolb,
Nucl.\ Phys.\ A {\bf 702}, 269 (2002).


\bibitem{Heiselberg:1995sh}
H.~Heiselberg and X.~N.~Wang,
Phys.\ Rev.\ C {\bf 53}, 1892 (1996).

\bibitem{Heinz:2002rs}
U.~Heinz and S.~M.~Wong,
Phys.\ Rev.\ C {\bf 66}, 014907 (2002).

\bibitem{Krasnitz:2002ng}
A.~Krasnitz, Y.~Nara and R.~Venugopalan,
Phys.\ Lett.\ B {\bf 554}, 21 (2003).

\bibitem{Kovchegov:2002nf}
Y.~V.~Kovchegov and K.~L.~Tuchin,
Nucl.\ Phys.\ A {\bf 708}, 413 (2002).

\bibitem{Mrowczynski:qm}
St. Mr\' owczy\' nski,
Phys.\ Lett.\ B {\bf 314}, 118 (1993).

\bibitem{Mrowczynski:xv}
St. Mr\' owczy\' nski,
Phys.\ Rev.\ C {\bf 49}, 2191 (1994).

\bibitem{Mrowczynski:1996vh}
St. Mr\' owczy\' nski,
Phys.\ Lett.\ B {\bf 393}, 26 (1997).

\bibitem{Shuryak:2002an}
E.~Shuryak,
arXiv:hep-ph/0205031.

\bibitem{Borghini:2000sa}
N.~Borghini, P.~M.~Dinh and J.~Y.~Ollitrault,
Phys.\ Rev.\ C {\bf 63}, 054906 (2001).

\bibitem{Borghini:2001vi}
N.~Borghini, P.~M.~Dinh and J.~Y.~Ollitrault,
Phys.\ Rev.\ C {\bf 64}, 054901 (2001).

\bibitem{Voloshin:1994mz}
S.~Voloshin and Y.~Zhang,
Z.\ Phys.\ C {\bf 70}, 665 (1996).

\bibitem{Poskanzer:1998yz}
A.~M.~Poskanzer and S.~A.~Voloshin,
Phys.\ Rev.\ C {\bf 58}, 1671 (1998).

\bibitem{Shuryak:2001me}
E.~V.~Shuryak,
Phys.\ Rev.\ C {\bf 66}, 027902 (2002).

\bibitem{Shuryak:2002qz}
E.~Shuryak and I.~Zahed,
Phys.\ Rev.\ D {\bf 67}, 014006 (2003).

\bibitem{Janik:2002nk}
R.~A.~Janik, E.~Shuryak and I.~Zahed,
Phys.\ Rev.\ D {\bf 67}, 014005 (2003).

\bibitem{fluct_hif} E.~Shuryak, Talk at CERN Workshop on Event-by-Event 
Fluctuations in Heavy-Ion Collisions, June 2001.

\bibitem{Wei59} E.S.~Weibel, Phys. Rev. Lett. {\bf 2} (1959) 83.

\bibitem{Mrowczynski:1999vi}
St. Mr\' owczy\' nski,
Acta Phys.\ Polon.\ B {\bf 31}, 2065 (2000).

\bibitem{Gazdzicki:ri}
M.~Gazdzicki and St. Mr\' owczy\' nski,
Z.\ Phys.\ C {\bf 54}, 127 (1992).

\bibitem{Alexopoulos:ft}
T.~Alexopoulos {\it et al.}  [E735 Collaboration],
Phys.\ Lett.\ B {\bf 353}, 155 (1995).

\bibitem{Brandt:2002qr}
A.~Brandt {\it et al.}  [UA8 Collaboration],
Eur.\ Phys.\ J.\ C {\bf 25}, 361 (2002).

\bibitem{Ostrovsky:2002cg}
D.~M.~Ostrovsky, G.~W.~Carter and E.~V.~Shuryak,
Phys.\ Rev.\ D {\bf 66}, 036004 (2002).


\end{thebibliography}
\end{document}